\documentclass{aa}
\usepackage{graphicx}
\usepackage{aalongtable}
\def\lunits{$\rm erg~s^{-1}$}
\def\funits{$\rm erg~cm^{-2}~s^{-1}$}
\def\cunits{$\rm cm^{-2}$}

\def\chandra{{\it Chandra~}}

\begin{document}
 \title{Chandra and Spitzer observations of CDFS X-ray obscured  QSOs}


  \titlerunning{X-ray Obscured QSOs}
    \authorrunning{I. Georgantopoulos et al.}

   \author{I. Georgantopoulos\inst{1}
  A. Georgakakis \inst{2}
          A. Akylas \inst{1}
          }

   \offprints{I. Georgantopoulos, \email{ig@astro.noa.gr}}

   \institute{Institute of Astronomy \& Astrophysics,
              National Observatory of Athens, 
 	      Palaia Penteli, 15236, Athens, Greece \\
              \and
         Astrophysics Group, Blackett Laboratory, Imperial College,  
       Prince Consort Road, SW7 2BZ, U.K. \\
             }

   \date{Received ; accepted }

   \abstract{We present \chandra and {\it Spitzer} data of the 186,
 extragalactic, hard 2-10 keV X-ray selected sources, 
 which lie in the central part of the Chandra Deep Field 
 South (CDFS). For the vast majority of sources 
 (99.5\%) there is a spectroscopic or photometric redshift available. 
 We classify 17 sources as X-ray obscured QSOs, according 
 to strictly X-ray criteria, i.e. defined as 
 having large hydrogen column densities ($N_H>10^{22}$ \cunits)
 and luminosities ($L_x>10^{44}$ \lunits).
 The surface density of X-ray obscured QSOs is $\sim$210 $\rm deg^{-2}$.  
 We find 18 candidate Compton thick $N_H>10^{24}$ \cunits sources,
 of which three have QSO luminosities ($L_x>10^{44}$ \lunits). 
 The X-ray obscured QSO comprise a mixed bag of objects,
 covering the redshift range $z=1.3-4.3$.    
 Eight of these show narrow line optical spectra,
 two show no obscuration in their optical spectra 
 presenting Broad Lines, while for the other seven there 
 is only a photometric redshift available.
 About half of the X-ray obscured QSOs show high X-ray to optical flux ratios,
 $X/O>1$, and red colours, I-3.6$\mu m>4$.
 Combination of the X-ray with the mid-IR $8\mu m$ or 24$\mu m$ flux can be used 
 as an additional diagnostic to sift out 
 the heavily obscured AGN. All X-ray selected QSOs present red mid-IR colours 
 and can be easily separated among mid-IR sources, demonstrating 
 that mid-IR selection provides a powerful tool for the detection 
 of obscured QSOs.  
         }

   \maketitle
%

\section{Introduction}

 Deep \chandra surveys have resolved a substantial fraction of the 
 X-ray background in the 2-10 keV band (Brandt \& Hasinger 2005). 
 The vast majority of the detected sources in these fields are
 AGN, both unobscured  ($\rm N_H<10^{22}$ \cunits) and  obscured
 ($\rm N_H>10^{22}$ \cunits),  with the latter dominating at faint
 fluxes. For example, at the source detection limit of the  Chandra
 Deep Field South (CDF-S), $f_X(\rm 2-10\,keV)\sim
 10^{-16}\rm \, erg \, s^{-1} \, cm^{-2}$, about 80\% of the sources
 have $\rm N_H>10^{22}$\,\cunits (e.g. Alexander et al. 2003; Akylas
 et al. 2006).

 However, there is a clearly a scarcity  of NL QSO.
 Steffen et al. (2003) find that the number
 of narrow line (NL) AGN decreases at bright luminosities and high
 redshift. Only a limited number of X-ray selected NL QSO
 ($L_X>10^{44}$\,\lunits) have been
 identified at high-$z$ (e.g. Stern et al. 2002, Norman et al. 2002). 

 These NL QSOs could be rare just because of selection
 effects (e.g. Treister et al. 2004). For example, rest-frame hard
 X-ray photons at high redshift can penetrate large obscuring columns
 but the observed optical emission, probing the rest-frame UV, will be
 easily diminished by a even small amount of dust. Such sources will
 therefore be optically faint hampering detailed analysis and may be
 overlooked in follow-up studies. These sources however, are expected
 to have high X-ray to optical flux ratio ($\log(f_x/f_o)>1$ or
 X/O) or very red optical/near-IR colours. Fiore et al. (2003) indeed,
 find that a fraction of the high X/O sources in the HELLAS2XMM survey
 are associated with NL AGN in agreement with the above
 scenario. Similarly, Brusa et al. (2005) find that a significant
 fraction of obscured X-ray sources are associated with Extremely Red 
 Objects, ERO, defined as having $R-K>5$, again supporting that
 optical selection effects may play an important role. 
    
 Alternatively, the scarcity of NL QSOs may suggest that the AGN
 unification model does not hold well at high luminosities. 
 It has been recently found that the fraction of X-ray obscured AGN 
 decreases with increasing luminosity 
 (Ueda et al. 2003), La Franca et al. 2005, Akylas et al. 2006) 
The physical interpretation could be that the highly
 luminous AGN blow away the obscuring screen or they photoionize the  
 surrounding gas. 

 In order to test the two different interpretations above, it is
 important to constrain the surface density of X-ray obscured QSOs 
 (defined here as X-ray luminous,  obscured sources) and
 to determine their properties. For instance, do X-ray obscured QSOs 
   have high X/O ratios? Can
 they be selected  through their red (e.g. R-K) colours? Do they all  
 have NL or a fraction presents BL optical spectra (e.g. Akylas et
 al. 2004).   

 Padovani et al. (2004) searched for X-ray obscured QSOs in the CDF-N and
 CDF-S surveys using a high luminosity  
 and a hardness ratio criterion, finding a few tens candidate X-ray obscured QSOs. 
 However, a number of the X-ray obscured  QSO redshifts and luminosities come 
 indirectly from the empirical X/O correlation  with X-ray luminosity 
 (Fiore et al. 2003, Barger et al. 2003).  Here, we  attempt to study 
 the properties of X-ray obscured QSOs in the hard X-ray selected (2-10\,keV)
 CDF-S sample.  A great advantage of this sample is that for the vast
 majority of the sources (except one out of  247) 
 there is either a spectroscopic or a
 photometric redshift available. This means that optical selection effects
 have little impact on our study. Moreover, we determine the X-ray
 spectral properties of the sources using spectral fittings instead of
 the more crude and indirect method of hardness ratios, minimizing any  
 uncertainties in the determination of the X-ray absorbing column density.  
 Additionally, the wealth of data that has been recently accumulated
 in the CDFS region as part of the GOODS survey (e.g. Hubble ACS,
 {\it Spitzer}, Very Large Telescope photometry and spectroscopy)    
 allows us to study in detail the properties of the X-ray obscured QSOs.

  We use throughout the paper 
 $\rm H_o=75 km~s^{-1}~Mpc^{-1}$, $\Omega_m=0.3, \Omega_\Lambda=0.7$.

\section{The Data}

 \subsection{X-ray data} 
The 1Ms CDFS data consist of 11 individual \chandra 
 (Advanced CCD imaging Spectrometer) ACIS-I 
 pointings with aim points separated by a few arcseconds.
 The aim point coordinates are $\alpha=3^h32^m28^s.0, \delta=-27^\circ
 48^{\prime}30^{\prime\prime}$ (J2000).
 More details are presented in Rosati et al. (2002) and 
 Giacconi et al. (2002). 247 sources are detected in the 
 2-10 keV band down to a flux limit of $1\times 10^{-16}$ \funits
 ($\Gamma=1.4$).   
 As the roll angles of the individual pointings are different,
 a fraction of the sources at the edge of the field-of-view 
  are detected only in a smaller number of pointings. 
 We choose here to  analyse only the sources which lie in 
all 11 pointings, in order to maximise the photon statistics. 
 Essentially our source selection is such that 
 the central field-of-view is covered. 
 188 sources are detected within all 11 pointings covering 
 an area of $\rm \sim0.05 deg^2$. Two sources are associated with  
 stars (Szokoly et al. 2004). Only one source has not been optically 
 identified and thus there is no redshift (photometric 
 or spectroscopic) available.

\begin{figure}
\includegraphics[width=8.0cm]{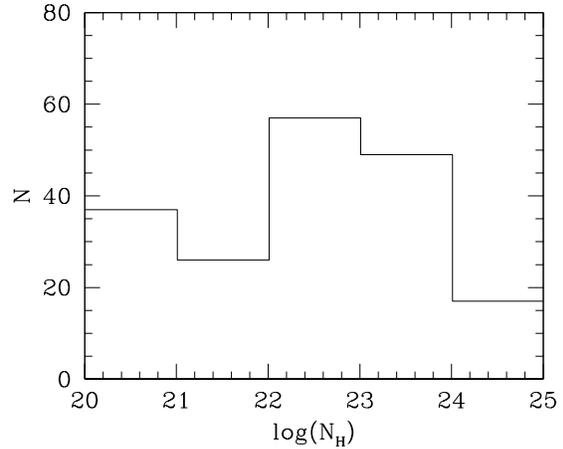}
\caption{The distribution of the rest-frame column density 
for all sources}
\label{nh}
\end{figure}

We use the {\sl PSEXTRACT} script in  the {\sl CIAO} v3.2  
 software package to extract spectra. There are 30 sources with more than 
 approximately 500 counts. The data are grouped so that there are 
  20 counts per bin for these sources and thus $\chi^2$
 statistics can apply. For the other sources which have 
 more limited photon 
 statistics we use the C-statistic technique (Cash 1979) 
 specifically developed to extract spectral information
 from data with low signal-to-noise ratio.
 We use the {\sl XSPEC} v11.2 software package for the spectral fits.  
 We fit the data using a power-law model absorbed by two 
 cold absorbers: wa*wa*po in {\sl XSPEC} notation.
 The first column is fixed to the Galactic  
 ($8\times10^{19}$ \cunits) while the second one is the 
 observer's frame {\it intrinsic} column density.
 Then the rest-frame column density scales as $(1+z)^{2.7}$ 
 (e.g. Barger et al. 2003). 
 In the case of the sources with limited 
 photon statistics ($<500$ counts), the power-law photon index 
 has been fixed to $\Gamma=1.8$.  
 The intrinsic luminosities are estimated using a  
  K-correction appropriate for the best value of $\Gamma$. 
 The X-ray spectral fits are presented in table \ref{spectralfits}. 
 The distribution of the rest-frame column density is 
 shown in Fig. \ref{nh}. All candidate Compton thick 
 sources are plotted in the last $N_H$ bin.    

\subsection{Optical and mid-IR}  
 Part of the CDFS has been observed with the 
 Advanced Camera Surveys (ACS) onboard 
 the {\it Hubble} Space Telescope
 as part of the GOODS survey (Giavalisco et al. 2004).
  We use here the F775W filter data 
 (roughly equivalent to the I band).
 We cross-correlate the X-ray sources 
 with the ACS data using a radius of 
 3 arcsec. As the surface density of sources 
 increases at faint magnitudes reaching $\sim 10^5$ 
 $\rm deg^{-2}~0.5mag^{-1}$ at I=26, Kashikawa et al. 2004,
 one has to be cautious about the possibility of chance 
 coincidence in the faintest cases (e.g. CDFS-610).
 For the sources which do not have I-band photometry 
 available, we quote R magnitudes from Giacconi et al. (2002). 
  
 Part of the CDFS has been observed by the IR {\it Spitzer} mission 
 (Werner et al. 2004). 
 The {\it Spitzer} Infrared Array Camera, IRAC, (Fazio et al. 2004) 8$\mu m$ and 
 Multiband Imaging Photometer, MIPS, (Rieke et al. 2004)  $24 \mu m$ fluxes are 
 are derived from the flux calibrated, background subtracted 
 images provided in the {\it Spitzer} archive. 
\footnote{\tiny http://data.spitzer.caltech.edu/popular/goods/Documents/}
 We cross-correlate the {\it Spitzer} with the \chandra sources
 again using a radius of 3 arcsec. 

\section{Results}

 \subsection{The X-ray obscured QSOs} 
\begin{figure}
\includegraphics[width=8.0cm]{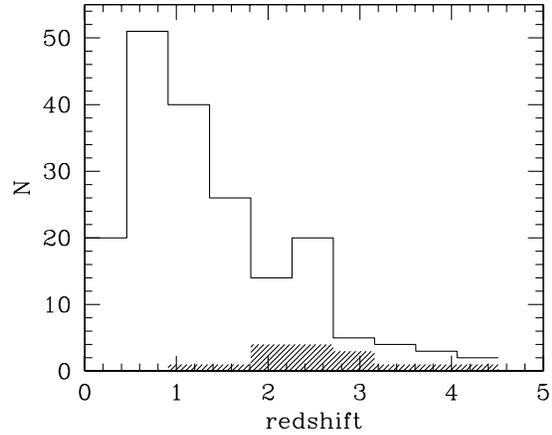}
\caption{The X-ray obscured QSO (hatched histogram) redshift distribution 
 compared with the total (open).}
\label{z}
\end{figure}

 23 sources are defined as  QSOs based 
 on their high intrinsic luminosity ($L_x>10^{44}$ \lunits). 
 Of these 18 (see table \ref{type-2}) present high absorbing column densities 
 ($N_H>10^{22}$ \cunits) and are thus classified as absorbed 
 QSOs (or X-ray obscured QSOs) purely on the basis of their 
 X-ray spectrum and luminosity. 
 However, at least one source (a BL QSO) is a borderline X-ray obscured QSO 
 as its column density uncertainty is relatively large.  
 Source CDFS-24,  which has  a BL optical spectrum, is consistent
 with $N_H=0$ \cunits  at the 90 \% confidence level. 
 We caution that a small error in the 
 measured column density at the  observer's frame
 (owing e.g. to a background fluctuation)  
 may translate to an erroneously high column density at the high redshifts
 probed here (see Akylas et al. 2006). 
 For example, as the rest-frame column density scales as $(1+z)^{2.7}$, a column 
 as low as  $\sim 2.5\times10^{20}$  at the observer's frame will  
 pass the $10^{22}$ \cunits rest-frame column threshold 
 at a redshift of z=3. 
 Hereafter, we exclude CDFS-24 from the X-ray obscured source sample.
 The resulting sample contains 17 sources.  
 Eight of the 17 X-ray obscured QSOs present NL optical spectra while 
 two sources are associated with BL AGN. For seven sources there 
 are no spectra available.    
 The redshift distribution is given in Fig. \ref{z}.   
 All sources apart from one lie at redshift $z>2$.

 We estimate the surface density of X-ray obscured QSOs 
  by taking into account the area covered by the 
 survey at a given flux. We estimate a 
 surface density of 210$\rm deg^{-2}$ 
 down to a flux of $f_{2-10}\approx 1.6\times10^{-15}$ \funits.  
 Five more luminous ($L_x>10^{44}$ \lunits) sources present no absorption 
 ($N_H>10^{22}$ \cunits) and are thus classified here as X-ray unobscured 
 QSOs (see table \ref{type-1}).
 The surface density of X-ray unobscured QSOs is $\sim 60$ $\rm deg^{-2}$
 at a flux of $f_{2-10}\sim 5\times 10^{-15}$ \funits.  
 The ratio of X-ray obscured to unobscured QSOs is R=3.4$\pm$1.7. However,
 at the same flux limit, $\sim 5\times 10^{-15}$ \funits, we estimate $R\sim1$. 
 Four of the X-ray unobscured QSOs  are associated with BL AGN, while for 
 the other one there is no spectrum available.

\subsection{Compton thick sources}
\begin{figure*}
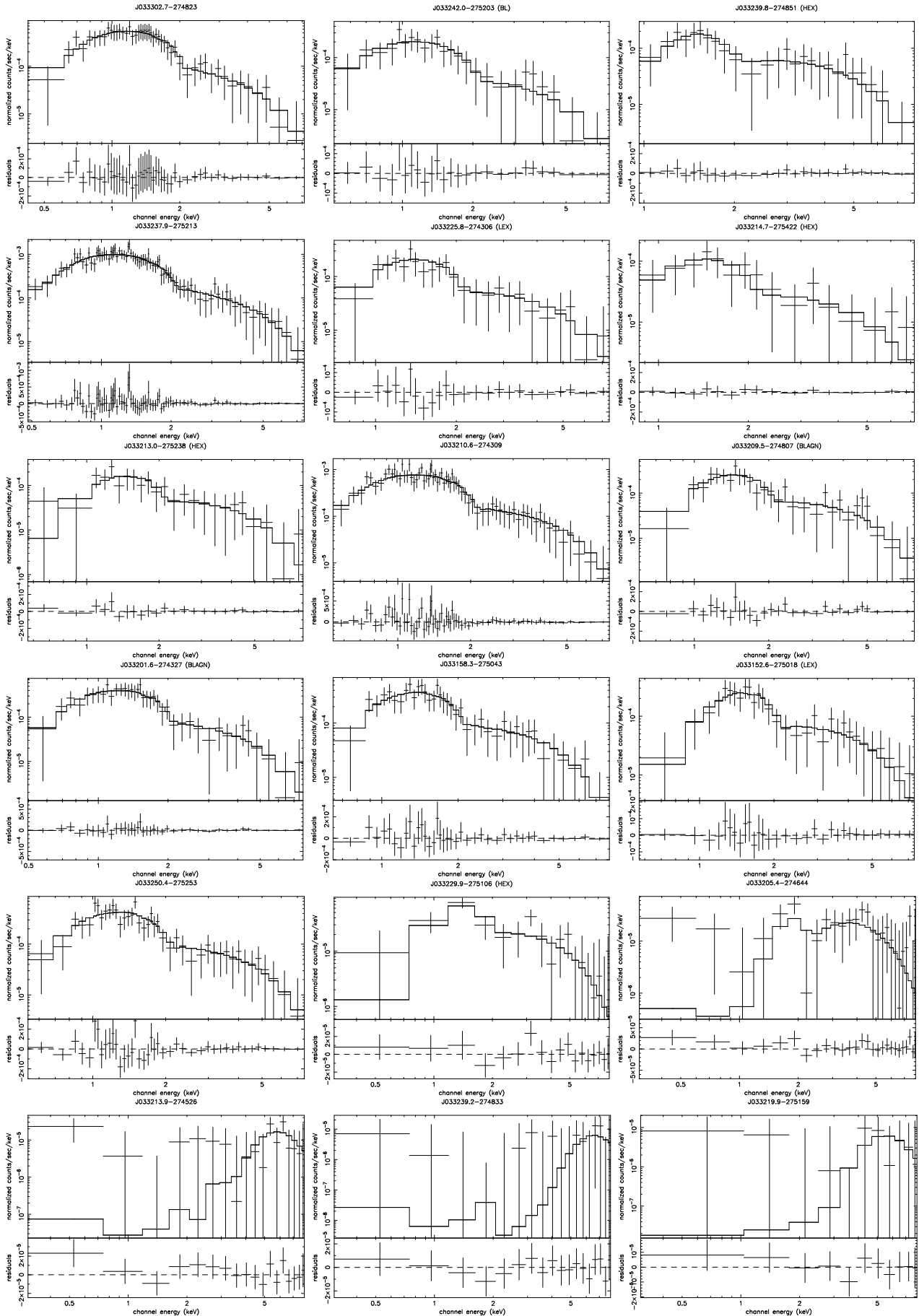

\rotatebox{270}{\includegraphics[width=4.cm]{178.eps}}
\rotatebox{270}{\includegraphics[width=4.cm]{141.eps}}
\rotatebox{270}{\includegraphics[width=4.cm]{137.eps}}\hfill \\
\rotatebox{270}{\includegraphics[width=4.cm]{129.eps}}
\rotatebox{270}{\includegraphics[width=4.cm]{93.eps}} 
\rotatebox{270}{\includegraphics[width=4.cm]{54.eps}}\hfill \\ 
\rotatebox{270}{\includegraphics[width=4.cm]{45.eps}}
\rotatebox{270}{\includegraphics[width=4.cm]{34.eps}}
\rotatebox{270}{\includegraphics[width=4.cm]{31.eps}}\hfill \\ 
\rotatebox{270}{\includegraphics[width=4.cm]{11.eps}}
\rotatebox{270}{\includegraphics[width=4.cm]{7.eps}}
\rotatebox{270}{\includegraphics[width=4.cm]{1.eps}}\hfill \\
\rotatebox{270}{\includegraphics[width=4.cm]{161.eps}}
\rotatebox{270}{\includegraphics[width=4.cm]{105.eps}}
\rotatebox{270}{\includegraphics[width=4.cm]{21.eps}}\hfill \\ 
\rotatebox{270}{\includegraphics[width=4.cm]{49.eps}}
\rotatebox{270}{\includegraphics[width=4.cm]{136.eps}} 
\rotatebox{270}{\includegraphics[width=4.cm]{72.eps}} \hfill \\
\caption{Power-law fits and $\chi^2$ residuals 
 to the spectra of the X-ray obscured QSOs}
 \label{fits}
 \end{figure*}

 The X-ray spectral fits for the X-ray obscured QSOs 
 are presented in Fig. \ref{fits}. 
 The X-ray spectral fits show that three sources apparently present very 
 large amounts of rest-frame absorption ($N_H>10^{24}$ \cunits) and
 are thus candidate Compton thick sources. 
 These are CDFS-600, 610 and 605 at redshifts
 1.327, 2.04 and 4.29 respectively.
 We caution that the absorption model we use 
 ({\sl WA} in {\sl XSPEC}) is only valid at column densities 
 lower than $10^{24}$ \cunits where photoelectric 
 absorption dominates the opacity. 
 At higher column densities Compton scattering 
  contributes significantly to the opacity.
 In these cases we have used the {\sl PLCABS} 
 model (Yaqoob 1997) which takes into 
 account multiple Compton scattering and 
 is valid for columns up to $5\times 10^{24}$ \cunits. 
 For higher columns we cannot obtain an accurate 
 enough estimate of the column but nevertheless 
 we can be confident that the source is Compton thick. 
  CDFS-600 is associated with a NL AGN  
 while the other two sources have only a photometric 
 redshift available.
 Another uncertainty in the determination 
 of the column density comes from the photometric 
 redshifts themselves.  
 Rigby et al. (2005) point out 
 that there might be an ambiguity on the photometric redshifts at least 
 in the case of optically faint sources.   
 They present {\it Spitzer} photometric observations 
 of 20 optically faint sources in the CDFS.   
 3 of their sources coincide with X-ray obscured QSOs here: 
 27, 45 and 159.  Rigby et al. (2005) derive 
 photometric redshifts using the additional 
 {\it Spitzer} bands. In the case of  e.g. CDFS-159 they find 
 a redshift discrepant with that  of Zheng et al. (2004) by $\delta z>1$.     
 In the case of Compton thick sources a strong 
 6.4 keV FeK line (equivalent width $\sim 1$ keV) is 
 usually observed (Matt et al. 1996). Unfortunately, 
 the spectra of the three candidate Compton thick 
 sources have limited statistics revealing  
 no evidence for an Fe line. The 90\% upper limits on the 
 equivalent width of the 6.4 keV line are as high as 
 several keV in all three cases.  

 In total we find 8 Compton thick AGN when we consider all 
 X-ray luminosities .
 These Compton thick AGN are revealed directly through  
 the detection of their absorption turnover . Such column 
 densities cannot be detected at low redshift 
 in the \chandra energy passband as the absorption 
 turnover occurs at energies $>10$ keV.  
 At high redshift the k-correction shifts
 the absorption turnover at low energies.   
 It is likely that these 8 AGN represent only a fraction of 
 the Compton thick sources in our sample. 
 Indeed, at lower redshifts or higher column densities 
 the column density cannot be directly detected. 
 Instead, the spectrum will appear flat. 
 We have thus looked for cases where leaving the 
 spectral index free, results in  
 a very flat spectrum: $\Gamma \sim1$ or flatter with    
 $\Delta C=2.7$. We find ten such cases,
 raising the total number of candidate Compton thick sources 
 to 18 (see table \ref{spectralfits}).

 \subsection{X/O ratio and optical-IR colour} 
  The optical (I or R)  magnitudes and X-ray to optical flux ratios, X/O, 
 are presented in table \ref{type-2}.
 Four X-ray obscured QSOs fall outside the ACS survey. For 
 these we present R  magnitudes instead from VLT/FORS.  
 For CDFS-72 there is only a magnitude lower limit as 
 this was not  detected at the limit of the FORS observations ($R=26$).  
 The  X/O is defined as the ratio of the 2-10 keV flux to 
 optical (I band) flux: 

 \begin{equation}
  X/O=  log(f_x/f_o) = 5.7+logf_x+I/2.5
 \end{equation} 
 
\begin{figure}
\includegraphics[width=8.0cm]{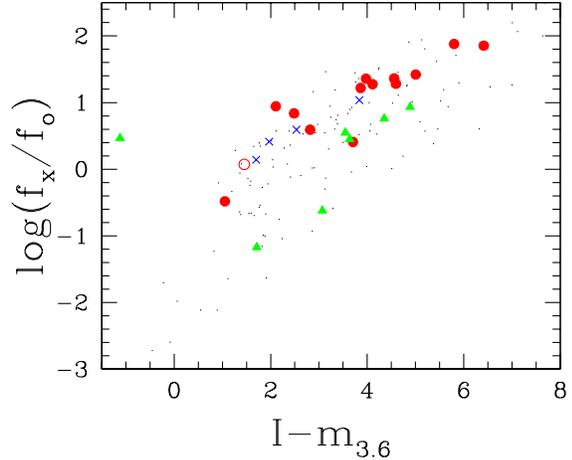}
\caption{The logarithm of the X-ray to optical flux ratio or  
 X/O as a function of the $I-3.6\mu m$
 colour. X-ray obscured and unobscured QSOs are denoted 
 with filled circles and crosses respectively. The open circle 
 denotes the border-line X-ray obscured QSO CDFS-24.
 The filled triangles represent the Compton thick AGN 
 with luminosity $\rm L_x<10^{44}$ \lunits. 
 The remaining sources are shown as dots.     
 }
\label{xo}
\end{figure}

 In the case where we have only R magnitudes available the X/O
 ratio is estimated using the equation in Hornschemeier et al. (2003).
 Three sources  (CDFS-6, 72, 76) have not been covered 
 by the {\it Spitzer} IRAC survey. Five sources have not been covered by 
 MIPS (CDFS-6, 72, 76, 61, 54).   
 All X-ray obscured QSOs which lie within the IRAC field-of-view have been detected   
 at $8\mu m$. Instead only seven have been detected by MIPS at $24 \mu m$. 
 Finally, the I-$3.6\mu m$ colour is given in Table \ref{type-2}. 
 The 3.6 $\mu m$ magnitude is estimated from the relation: 
 
 \begin{equation}
  m_{3.6 \mu m}= -2.5~logf_{3.6\mu m}  + 23.9  
 \end{equation} 

 where the flux $f_{3.6}$ is in units of $\rm \mu Jy$. 
The X/O ratio as a function of the I-3.6$\mu m$ colour is given 
 in Fig. \ref{xo}. There is a strong correlation between 
 the X/O and the I-3.6$\mu m$ colour in the sense that the 
 redder sources present high X/O. All the high X/O sources 
 (X/O$>$1 or Extreme X/O sources EXO) have 
  I-3.6$\mu m > 4 $. This colour is roughly equivalent to an Extremely
 Red Object (ERO) defined as having $R-K>5$. However, not all X-ray obscured QSOs are 
   EXOs or EROs. Reversely, one  of the X-ray unobscured QSOs is an EXO (CDFS-67).  

 \subsection{Mid-IR properties} 
\begin{figure*}
\includegraphics[width=8.0cm]{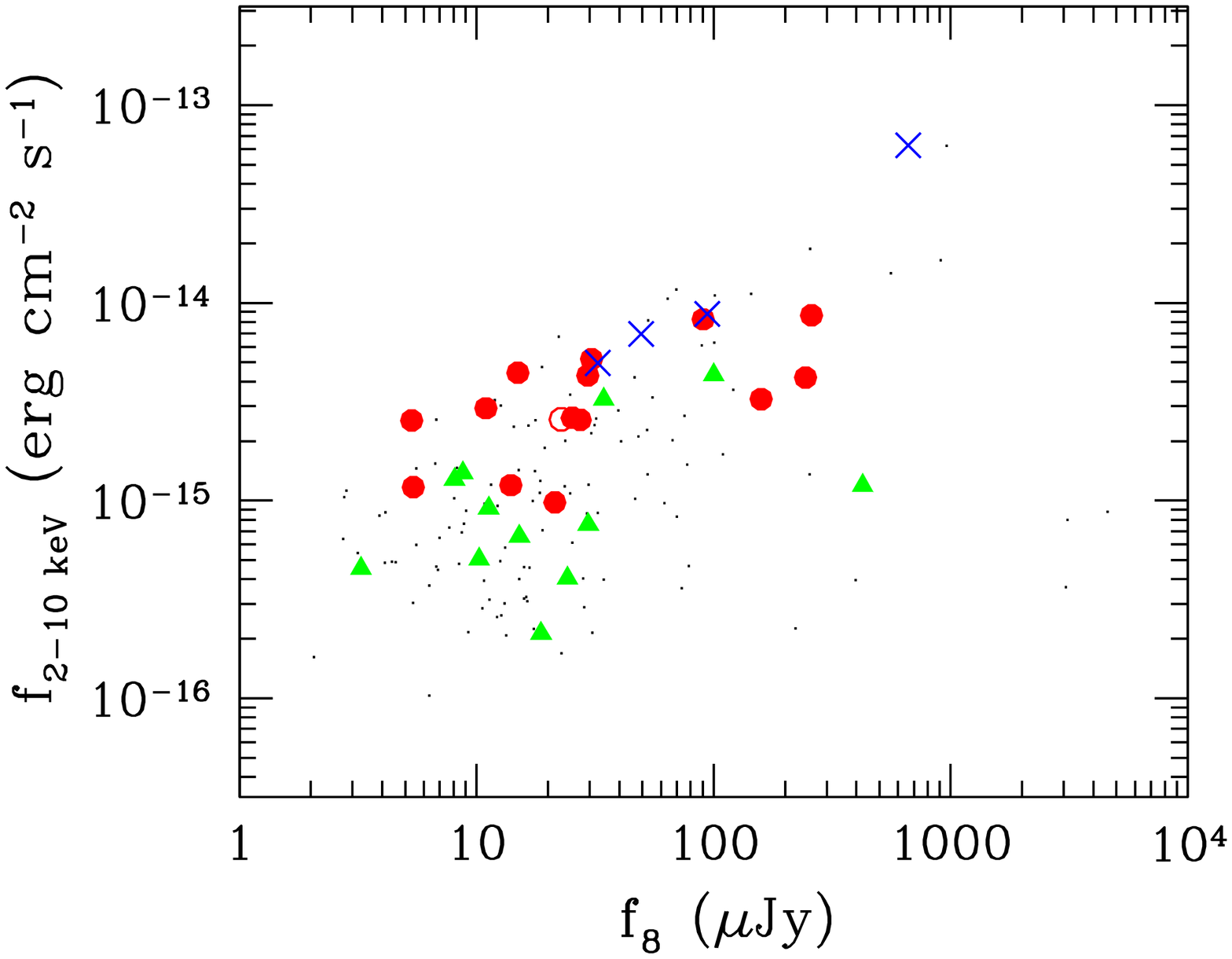} \includegraphics[width=8.0cm]{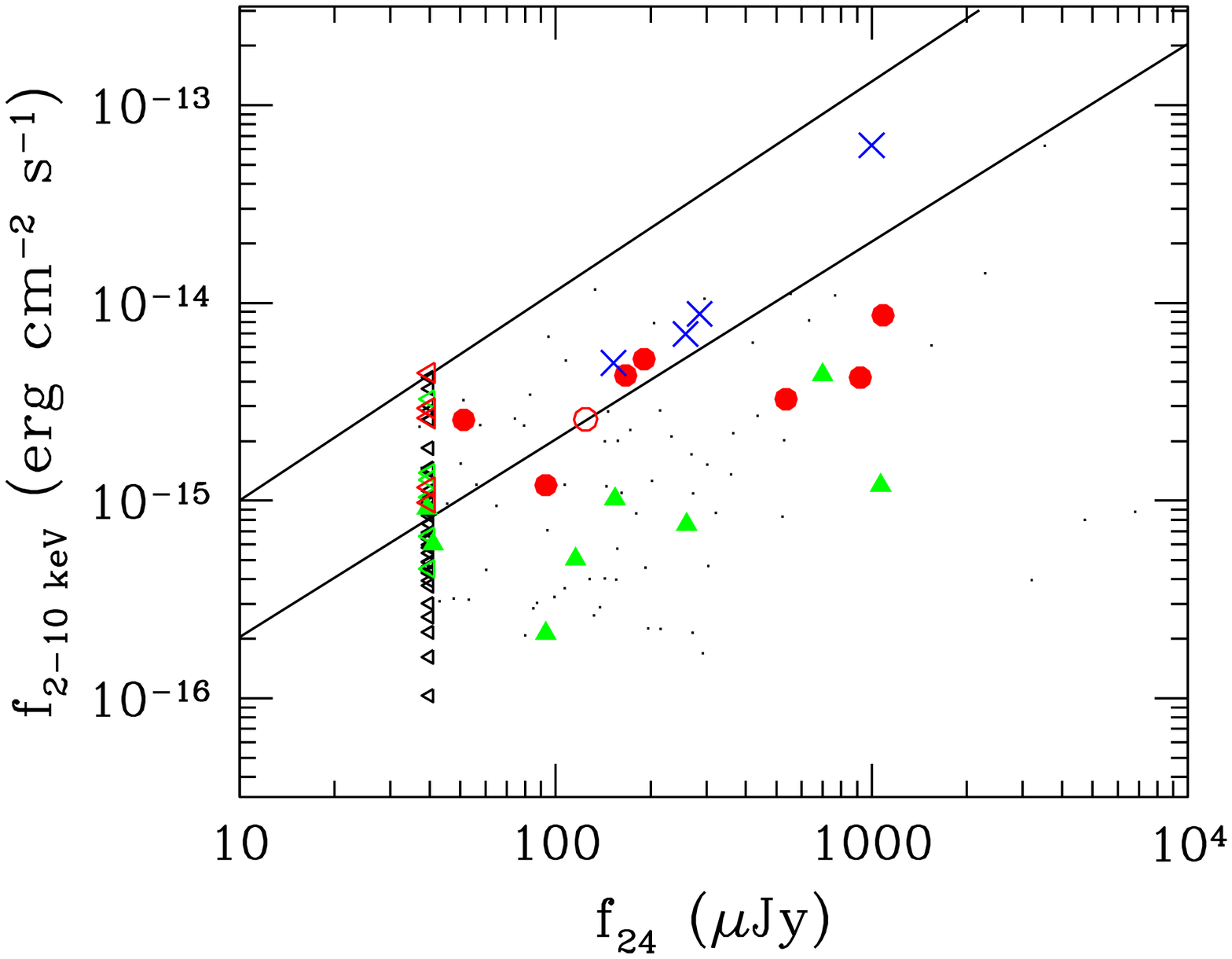} 
\caption{The 2-10 keV absorbed X-ray flux versus 
 the {\it Spitzer} IRAC 8$\mu m$ (left panel) 
 and MIPS 24$\mu m$ monochromatic flux 
 for hard X-ray selected sources in the CDFS.
 Symbols as in Fig. 3; 
 open triangles denote 24 $\mu$m luminosity upper limits.  
 The solid lines in the right panel denote  
 the region occupied by the Piccinotti AGN sample 
 (see Alonso-Herrero et al. 2004).} 
\label{xir}
\end{figure*}

 Additional clues on the nature of the X-ray obscured QSOs can be 
 provided by their mid-IR properties. 
 The X-ray to mid-IR flux ratios provide a useful diagnostic 
 for identifying obscured AGN. Unobscured AGN have ratios 
 around unity (e.g. Lutz et al. 2004). 
 In Fig. \ref{xir} we present the absorbed X-ray flux against the IRAC 
 $8\mu m $ (left panel) and $24\mu m$ (right panel) flux density 
 for all X-ray sources. 
 The solid lines (adapted from Alonso-Herrero et al. 2004) 
 denote the region which is occupied by 
 the hard X-ray selected AGN in the HEAO-1 sample of 
 Piccinotti et al. (1982).   
There are four X-ray obscured QSOs which lie 
 below the AGN locus of points  in the 24$\mu m$ diagram.
 The same X-ray obscured QSOs have low X-ray fluxes relative to 
 their $8\mu m$ flux densities.  
 These are associated with a BL QSO (CDFS-62),
 and three NL QSOs (CDFS-202, CDFS-45, CDFS-31). These have less 
 X-ray emission for their mid-IR flux density and 
 thus they should be associated with the most heavily 
 absorbed sources. Surprisingly, none of the  
 candidate Compton thick sources are among these. 
  The BL QSO CDFS-62 presents
 no evidence of reddening in the optical having $I-3.6\approx1$.
 In contrast, sources CDFS-45 and CDFS-31 have $I-3.6> 4$ being 
 among the redder sources in our sample.  
 CDFS-202 presents little reddening with $I-3.6=2.8$.  
 Norman et al. (2002) present in detail the properties 
 of this NL QSO at a redshift of 3.7. 
 
\begin{figure}
\includegraphics[width=8.0cm]{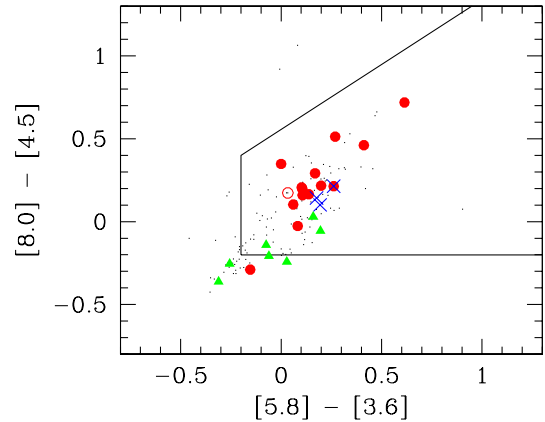}
\caption{ {\it Spitzer} IRAC colours. 
 We denote with solid lines the  region 
 which is used as an AGN diagnostic tool 
 in mid-IR according to Lacy et al. Symbols as in Fig. \ref{xo}.  
 }
\label{lacy}
\end{figure}

 In the mid-IR, AGN present red, featureless spectra (Hao et al. 2005).
 Thus the mid-IR colours can provide a powerful tool to identify 
 AGN (Lacy et al. 2004, Stern et al. 2004, Hatziminaoglou et al. 2005). 
 In Fig. \ref{lacy} we plot the mid-IR colours of our sample. 
 The solid lines denote the ``red'' region empirically defined by 
 Lacy et al. (2004) to contain 
 luminous AGN. Lacy et al. find that the normal galaxies lie in the blue part 
 of the diagram having colours clustering around [8.0]-[4.5]=-0.5
 and [5.8]-[3.6]=-0.5. The vast majority of our X-ray 
 selected of  QSOs are ``red''. The only source lying 
 outside (albeit only marginally) the red AGN region defined 
 by Lacy et al. is CDFS-600. 

\begin{table*}
\begin{center}
\caption{The X-ray obscured QSOs}
\label{type-2}
\begin{tabular}{lccccccccccc}
\hline
ID$^1$ & Name & $\rm N_H^2$ &    z$^3$ &   Opt. Mag.$^4$ &  Colour$^{5}$ & $L_x^6$
  & $f_{8\mu m}^7$ & $f_{24\mu m}^8$ & $fx^9$ &  ${X/O}^{10}$ & class$^{11}$       \\
\hline 
6 &  J033302.7-274823 & $4.0^{+1.6}_{-1.9}$ & 2.46 & 25.67$^a$ & 4.17$^a$ & 44.23 & - & - & -14.40 & 1.36 & - \\
24$^\dagger$ & J033242.0-275203 &$4.0^{+4.6}_{-4.0}$ & 3.610 & 22.40 & 1.45 & 44.42 & 22.7 & 124.3 & -14.59 & 0.07 & BL \\
27 & J033239.8-274851 &$47.8^{+8.6}_{-7.8}$ & 3.064 & 24.71 & 3.86 & 44.50 & 29.4 & 166.6 & -14.37 & 1.21 & NL \\
31 &J033237.9-275213&  $2.1^{+0.5}_{-0.6}$ & 1.603 & 24.11 & 4.59 & 44.17 & 259.1 & 1085.4 & -14.06 & 1.28 & NL \\
45 &  J033225.8-274306 &$12.2^{+3.3}_{-2.8}$ & 2.291 & 25.37 & 4.56 & 44.08 & 158.5 & 535.7 & -14.49 & 1.36 & NL \\
54 & J033214.7-275422 &$18.8 ^{+6.1}_{-5.5}$& 2.561 & 25.69$^a$ & - & 44.08 & 5.3 & - & -14.60 & 0.87 & NL \\
57 & J033213.0-275238 &$18.8^{+5.2}_{-4.6}$ & 2.562 & 24.17 & 2.48 & 44.15 & 10.9 & $<40$ & -14.53 & 0.84 & NL \\
61 & J033210.6-274309 & $2.4^{+0.7}_{-0.9}$ & 2.02 & 24.51 & 5.00 & 44.28 & 90.3 & -& -14.08 & 1.42 & - \\
62 & J033209.5-274807 &$23.2^{+4.5}_{-4.2}$ & 2.810 & 20.49 & 1.05 & 44.39 & 244. & $920.2$ & -14.38 & -0.48 & BL \\
68 & J033201.6-274327 & $6.9^{+3.0}_{-3.0}$ & 2.726 & 23.99 & 2.11 & 44.27 & 14.9 & $<40$ &  -14.35 & 0.94 & BL \\
72 & J033158.3-275043 & $10.4^{+3.1}_{-2.2}$ & 1.99 & $>$26$^a$ & - & 44.09 & - & - & -14.33 & $>1.56$ & - \\
76 & J033152.6-275018 & $19.8^{+3.8}_{-3.6}$ & 2.394 & 24.50$^a$ & 4.64$^a$ & 44.34 & - & - & -14.28 & 0.72 & NL \\
159 & J033250.4-275253 &$9.5^{+6.7}_{-3.8}$ & 3.30 & 24.86 & 3.97 & 44.65 & 30.6 & 190.5 & -14.28 & 1.35 & - \\
202 & J033229.9-275106 &$36.8^{+20.5}_{-15.7}$ & 3.700 & 24.54 & 2.82 & 44.12 & 1.0 & 93.1 & -14.92 & 0.59 & NL \\
227& J033205.4-274644 &$65.4^{+22.0}_{-16.5}$ & 2.18 & 26.90 & 5.80 & 44.00 & 25.3 & $<40$  & -14.58 & 1.87 & - \\
600$^\ast$& J033213.9-274526 &$380^{+180}_{-180}$ & 1.33 & 23.25 & 3.71 & 44.43 & 27.4 & 51.2 & -14.58 & 0.41 & NL \\
605$^\ast$& J033239.2-274833 & $2700^{+3800}_{-1700}$ & 4.29 & 26.27 & 4.11 & 44.95 & 5.4 & $<40$ & -14.92 & 1.27 & - \\
610$\ast$& J033219.9-275159 &$520^{+580}_{-280}$ & 2.04 & 27.90 & 6.41 & 44.18 & 21.4 & $<40$ & -15.01 & 1.85 & - \\
\hline
\multicolumn{7}{l}{$^1$ ID number from Giacconi et al. (2002)} \\
\multicolumn{7}{l}{$^2$ Rest-frame column density in units $10^{22}$ \cunits} \\
\multicolumn{7}{l}{$^3$ Redshift from: (a) Szokoly et al. (spectroscopic) or 
                         (b) Zheng et al. (photometric)} \\
\multicolumn{7}{l}{$^4$ Optical ACS I magnitude; (a) denotes R magnitude instead} \\
\multicolumn{7}{l}{$^5$ I-3.6 colour; (a) denotes R-K colour instead} \\
\multicolumn{7}{l}{$^6$ logarithm of 2-10 keV intrinsic luminosity (\lunits)} \\
\multicolumn{7}{l}{$^7$ {\it Spitzer} IRAC 8 $\mu m$ flux density ($\mu Jy$)} \\
\multicolumn{7}{l}{$^8$ {\it Spitzer} MIPS 24 $\mu m$ flux density ($\mu Jy$)} \\
\multicolumn{7}{l}{$^9$ absorbed 2-10 keV X-ray flux \funits as derived from the spectral fits } \\
\multicolumn{7}{l}{$^{10}$ logarithm of X-ray to optical flux ratio} \\
\multicolumn{7}{l}{$^{11}$ Optical Spectroscopic classification: 
  narrow line (NL), broad line (BL)} \\
\multicolumn{10}{l}{$^{\dagger}$ Border-Line X-ray Obscured QSO, consistent with
 $N_H<10^{22}$ \cunits at the 90\% confidence level}  \\
\multicolumn{10}{l}{$^{\ast}$ Compton thick sources; column densities have been 
 derived using the plcabs model} \\
\end{tabular}
\end{center}
\end{table*}

\begin{table*}
\begin{center}
\caption{X-ray unobscured QSOs}
\label{type-1}
\begin{tabular}{lccccccccccc}
\hline
ID$^1$ &  Name &   $\rm N_H^2$ &    z$^3$ &   Opt. Mag.$^4$ &  Colour$^{5}$ & $L_x^6$
  & $f_{8\mu m}^7$ & $f_{24\mu m}^8$ & $fx^9$ &  ${X/O}^{10}$ & class$^{11}$       \\
\hline 
11 & J033260.0-274748& $<0.2$ & 2.579 & 21.83$^a$ & 2.73$^a$ & 44.54 & - & -& -14.15 & 0.08 & BL   \\
22 & J033243.3-274915& $<0.2$ & 1.920 & 22.54 & 1.97 & 44.16 & 32.4 & 152.7 &-14.30 & 0.41 & BL \\
42 &J033227.1-274105&  $<0.02$ & 0.73 & 19.11 & 1.70 & 44.25 & 661.5 & 999.7 & -13.20 & 0.14 & - \\
60 & J033211.0-274415& $<0.10$ & 1.615 & 22.37 & 2.53 & 44.22 & 93.9 & 285.7 & -14.05 & 0.59 & BL \\
67 & J033202.5-274601& $<0.43$ & 1.616 & 23.73 & 3.84 & 44.06 & 49.4 & 258.0 & -14.16 & 1.03 & BL \\
\hline
\multicolumn{7}{l}{$^1$ ID number from Giacconi et al. (2002)} \\
\multicolumn{7}{l}{$^2$ Rest-frame column density in units $10^{22}$ \cunits} \\
\multicolumn{7}{l}{$^3$ Redshift from: (a) Szokoly et al. (spectroscopic) or 
                         (b) Zheng et al. (photometric)} \\
\multicolumn{7}{l}{$^4$ Optical ACS I magnitude; (a) denotes R magnitude instead} \\
\multicolumn{7}{l}{$^5$ I-3.6 colour; (a) denotes R-K colour instead} \\
\multicolumn{7}{l}{$^6$ logarithm of 2-10 keV intrinsic luminosity (\lunits)} \\
\multicolumn{7}{l}{$^7$ {\it Spitzer} IRAC 8 $\mu m$ flux density ($\mu Jy$)} \\
\multicolumn{7}{l}{$^8$ {\it Spitzer} MIPS 24 $\mu m$ flux density ($\mu Jy$)} \\
\multicolumn{7}{l}{$^9$ logarithm of absorbed 2-10 keV X-ray flux \funits as derived from the spectral fits} \\
\multicolumn{7}{l}{$^{10}$ logarithm of X-ray to optical flux ratio} \\
\multicolumn{7}{l}{$^{11}$ Optical Spectroscopic classification: 
  narrow line (NL), broad line (BL)} \\
\end{tabular}
\end{center}
\end{table*}

\section{Discussion}
 We have derived the X-ray spectral properties for all 186 hard 
 extragalactic (2-10 keV) X-ray selected sources within the central 
 field-of-view of the CDF-S. We define X-ray obscured QSOs, strictly using 
 X-ray criteria and regardless of their optical spectrum, as those 
 having high unobscured luminosities $L_x>10^{44}$ \lunits and high 
 obscuration column density $N_H>10^{22}$ \cunits.
 We find 17 X-ray obscured QSOs among the 186 sources. The ratio of 
 X-ray unobscured to X-ray obscured QSOs ($L_x>10^{44}$ \lunits)  is $3.4\pm 1.7$. 
 This consistent with  the ratio at lower luminosities ($L_x<10^{44}$ 
 \lunits) which is $R\approx2.5\pm 0.5$.
 At first this comes at odds with recent findings 
 which show that the fraction of obscured 
 AGN decreases significantly at high luminosities:  
 Ueda et al. (2003), La Franca et al. (2005), Akylas et al. (2006)
 find ratios, R, well below unity, at high luminosities. 
 This apparent discrepancy   
 could be attributed to the fact that we are using a pencil beam 
 flux limited survey  and thus we cannot probe large enough volumes to detect
 a high number of luminous unobscured QSOs (see also the discussion by  
 La Franca et al. 2005 and Akylas et al. 2006).

 The optical spectra of the X-ray obscured QSOs show that these comprise 
 a mixed bag of objects. Eight sources present NL spectra, two 
 are associated with BL AGN while for seven more there are no 
 spectra available. 
 The BL source CDFS-062 presents  a  
 high column, $2\times10^{23}$ \cunits. This  column should correspond 
 to $A_V\sim 100$ using the Galactic dust-to-gas ratio 
 (Bohlin et al. 1978) or equivalently to $A_{2000 \AA}\sim 300$ 
 mags in the UV (Richards et al. 2003).  In actuality CDF-062 has a broad 
 CIII] emission line implying that the reddening must be negligible.  
 Such peculiar BL AGN with high X-ray obscuration but 
 little optical reddening have often been  reported
 in the literature (e.g. Elvis et al. 1998, Maiolino et al. 2001,
 Georgantopoulos et al. 2003, Perola et al. 2004, Akylas et al. 2004).
 It is possible that some of the absorption may be intervening.   
 The most prominent and well studied case of BL QSOs with large amounts of X-ray 
absorption are  the Broad Absorption Line (BAL) QSOs. Although 
 the X-ray spectra of BAL are  usually absorbed 
 by X-ray column densities of $10^{23}$ \cunits 
 (eg  Gallagher et al. 2001) they present only a 
 little amount of optical reddening (Brotherton et al. 2001).
 The optical spectrum of CDF-062 presents prominent,
 blueshifted  absorption lines of CIV and SiIV
 with a FWHM of about 5.000 $\rm km~s^{-1}$.
 The absorption line widths are clearly less 
 than those typical for BAL QSOs. 
 This suggests that a fraction of the X-ray absorbed 
 BL QSOs may be associated with a population of 
 such ``mini-BAL'' QSOs. CDFS-24 and 68 also present  
 evidene for blueshifted absorption lines.  
 
A large fraction of our X-ray obscured QSOs have high X/O ratios. 
Fiore et al. (2003) postulate that this is because at high redshift 
the nuclear emission  is obliterated by dust. The same high X/O  sources  
present very red colours having I-3.6$\mu m > 4$ and can be classified 
as EROs. Indeed it has been shown that X-ray detected EROs  have 
high X/O ratios (Lehmann et al. 2001, Alexander et al. 2001, 
Brusa et al. 2005).  It is likely that these EROs  have red colours 
because they are associated with elliptical galaxies at high redshift. 
The high X/O ratios can be explained as, on one hand the optical 
nuclear light is attenuated, while on the other hand 
 the X-ray emission emerges 
relatively unscathed.  Despite the fact that there is a very strong 
correlation between the X/O ratio and the I-3.6 colour there are some 
EROs which have X/O ratios characteristic of normal AGN, X/O$\sim0$, 
(see also Koekomoer et al. 2004). It is important to stress that not 
all X-ray obscured QSOs are red or high X/O sources. For example the X-ray obscured QSO 
CDFS-202 (Norman et al. 2002) has bluish colours with I-3.6=2.8, or R-K=2.5. 
Norman et al. (2002) assert that the blue colour is because of a strong 
emission line in the R filter. Nevertheless, the I-3.6 colour of CDFS-202    
 is blue as well. 

All X-ray obscured QSOs which have been observed by IRAC, 
 have been detected at $8\mu m$. In contrast, less than half have been 
 detected at 24$\mu m$. The X-ray to mid-IR  diagram can 
 provide additional   
 information on which sources have large intrinsic columns: 
 an absorbed source should have a low ratio of absorbed to mid-IR
 flux provided that the latter is an isotropic indicator of 
  the nuclear emission.
 Only four sources appear to be absorbed according to the above criterion.
 Then, by comparison, the X/O ratio or the I-3.6 colour provide a rather more 
 efficient diagnostic for separating X-ray obscured AGN. 
 This may be attributed to the high redshifts of our sources. 
 It is argued by Fadda et al. (2002) that  at high redshift, the X-ray 
 to mid-IR flux may increase rapidly, as the 
 IR flux corresponds to shorter wavelengths while the 
 X-ray flux is less affected by photoelectric absorption.
 Nevertheless, the X-ray to mid-IR flux ratio may provide 
 a tool complementary to the X/O ratio and I-3.6 colour. 
 Indeed, among the four sources with low X-ray to mid-IR ratio,
 CDFS-62 and CDF-202 have neither red colour nor a high X/O ratio.

Lacy et al. (2004) have demonstrated that the mid-IR colours can be 
efficiently used to separate AGN from normal galaxies. In particular 
Lacy et al. argue that the AGN have relatively red mid-IR colours (5.8-3.6 
against 8.0-4.5 $\mu m$) compared with normal galaxies. Barmby et al. 
(2006) argue that the X-ray selected AGN in the Extended 
Groth Strip Survey (EGS) occupy both blue and red mid-IR colours and hence such
 colours cannot be used efficiently to separate them from  normal galaxies. 
In our case we find that the mid-IR diagnostics used by Lacy et al. provide  a 
very powerful diagnostic for the selection of X-ray obscured QSOs. 
 Only one of our  X-ray obscured QSOs
 lies outside (albeit marginally) the AGN region. The discrepancy with the
 findings of Barmby et al. probably arises because   we are dealing only
 with luminous AGN. The EGS sources of Barmby et al. span a large range of 
luminosities: hence it is possible that the mid-IR colours of the less 
luminous sources (Seyfert-like luminosities) are contaminated by the host galaxy.

 The X-ray spectral fits show that three  of the X-ray obscured QSOs are
 Compton thick sources. Two of the Compton thick sources 
 present high X/O and red colour.  
 The number of Compton thick sources amounts to 18 
(or about 10 \% of all sources ) if we consider all X-ray luminosities. 
 This figure is roughly consistent with
 the predictions of Comastri et al. (2001) for the number of Compton thick
 sources at these flux levels (about 7\% at $5\times10^{-16}$ \funits).
 The predictions of Comastri et al. (2001) 
 are based on  the distribution of the column densities observed in the 
 local Universe by Risaliti et al. (1999).
  Our work increases substantially   
 the number of known Compton thick sources at high redshifts. 
 Other examples of Compton thick sources at high redshift 
 include the four sources at $z>2$ from the sub-mm selected sample 
 of Alexander et al. (2005) and the NL QSO at z=3.288 from the 
 Lynx field (Stern et al. 2002). 
 Moreover, two additional NL QSOs in the CDFS have been reported as Compton thick:
 CDFS-202 (Norman et al. 2002) and CDFS-263  (Mainieri et al. 2005).
 Our spectral fits place these sources marginally outside 
 the $N_H>10^{24}$ \cunits regime. Nevertheless, these are still consistent 
 with being Compton thick given the spectral fit uncertainties. 
 Moreover, the adopted slope of the photon index
 affects significantly the derived $N_H$.  
 Polletta et al. (2006) report the discovery of two
 Compton thick QSOs at redshifts z=2.54 and 2.43 in the Chandra/SWIRE 
 survey in the Lockman Hole. 

 Just before the submission of this paper, Tozzi et al. (2006) 
 presented an analysis of the X-ray spectral properties of all 321
 extragalactic  sources in the CDFS. Tozzi et al. find 14 Compton thick sources, 
 defined as having a reflection dominated spectrum.
 In particular, Tozzi et al. use the {\sl PEXRAV} model in XSPEC to fit the 
 spectra of these 14 sources. Interestingly, the Compton thick samples of Tozzi et al. 
 and ours have little overlap: only sources CDFS-153, 505, 531 and 610
 are common. 
 However, Tozzi et al. do not include in their Compton thick sample the sources 
 which present column densities $N_H>10^{24}$ \cunits, 
  i.e. CDFS-599, 600, 601, 605, 609, 634. 
 This would alleviate the discrepancy between the two samples.	   
  Finally, there are sources in their sample presenting very flat spectral 
 index, e.g. CDFS-25 with $\Gamma=0.29^{+0.21}_{-0.20}$, which should be 
 considered as candidate Compton thick sources.

\section{Conclusions} 

We have explored the X-ray optical and mid-IR properties of X-ray obscured QSOs on 
the CDFS.  We classify X-ray obscured QSOs on the basis of purely X-ray criteria 
i.e. high luminosity $\rm L_x>10^{44}$ \lunits 
 combined with high hydrogen column density $N_H>10^{22}$ \cunits. 
We are selecting X-ray obscured through the 186 hard extragalactic X-ray sources which
 lie in the central FOV of the CDFS. We have derived proper X-ray spectra
 (instead of hardness ratios) for all our sources. Photometric 
or spectroscopic redshifts exist for the vast majority of the sources 
(185 out of 186) ascertaining that optical selection biases are of no
 importance in our study. We find 17 X-ray obscured QSOs spanning the redshift 
 range 1.3-4.2.  Our main results can be summarised as follows:

1. The surface density of X-ray obscured QSOs is  210 $\rm deg^{-2}$ 
 at $f_{2-10}\sim 10^{-15}$ \funits. 

2. Our X-ray obscured QSO sample comprises of sources with both NL and BL spectra. 
The BL source with the highest absorption ($10^{23}$ \cunits) is a mini-BAL. 

3. Three out of  17 X-ray obscured QSOs present very high column densities 
 ($>10^{24}$ \cunits) and are classified as Compton thick QSOs. 
 When we consider all luminosities, we classify 18 sources as candidate 
 Compton thick AGN, based on the detection of either high column densities or 
 flat X-ray spectra. 

4. About half of our X-ray obscured QSOs present a very high X-ray 
 to optical flux ratio, 
X/O$>$1. These present very red colours (I-3.6 $\mu m$) and most can 
be classified as EROs.   

5. The X-ray to mid-IR flux ratio can provide a complementary  tool to  
select heavily obscured sources. Four out of 17 X-ray obscured QSOs 
 appear to have a low
 X-ray to mid-IR flux ratio; only two of these four sources 
 have either a red colour or a high X/O ratio. 

6. Practically all 17 QSOs have red mid-IR colours  and can be very 
easily separated from the normal galaxies detected in mid-IR as 
 pointed out by Lacy et al. (2004).
 This suggests that the mid-IR selection provides a powerful tool to 
 detect  obscured QSOs at high redshift.

\begin{acknowledgements}
 We are grateful to the anonymous referee for his 
 numerous corrections and suggestions. 
 We acknowledge use of ESO/GOODS data. 
 The Chandra data were taken from the Chandra Data Archive 
 at the Chandra X-ray Center.   
\end{acknowledgements}

\begin{longtable}{ccccccccccc}
\caption{\label{spectralfits} Spectral fits} \\
\hline \hline 
ID & Name$^1$ & $\Gamma^2$ & $N_H^3$ & $L_x^4$ & Class$^5$ & z$^6$ & flux$^7$ & $\chi^2$/d.o.f \\
\hline \hline
\endfirsthead
\caption{continued}\\
\hline\hline 
ID & Name$^1$ & $\Gamma^2$ & $N_H^3$ & $L_x^4$ & Class$^5$ & z$^6$ & flux$^7$ & $\chi^2$/d.o.f \\
\hline
\endhead
\hline
\endfoot 
3    & J033306.0-274651 & 1.80                    &       0.86$_{-0.29}^{+0.32}$ 	&41.52 &  Galaxy &  0.22 & 0.21 &  -   \\
4    & J033303.7-274519 & $1.93_{-0.15}^{+0.20}$  &       $<0.43$ 			&43.50 &    -    &  1.260 & 0.34 & 0.90 \\ 
6    & J033302.7-274823 & $2.25_{-0.19}^{+0.28}$  &       4.02$_{-1.88}^{+1.61}$ 	&44.23 &   -  &  2.46 & 0.39 & 0.63 \\ 
10   & J033259.9-274627 & 1.80                    &       2.71$_{-0.48}^{+0.53}$ 	&42.55 &   LEX   &  0.424 & 0.51 &  -   \\
11   & J033260.0-274748 & $1.97_{-0.09}^{+0.11}$  &       $<0.59$ 			&44.54 &  BLAGN  &  2.579 & 0.71 & 1.16 \\ 
12   & J033259.8-275031 & $2.20_{-0.18}^{+0.21}$  &       $<0.07$ 			&41.69 &   ABS   &  0.251 & 0.22 & 1.00 \\ 
15   & J033253.0-275120 & $1.91_{-0.22}^{+0.08}$  &       0.42$_{-0.42}^{+0.50}$ 	&43.51 &  BLAGN  &  1.23 & 0.37 & 1.11 \\ 
17   & J033249.3-275505 & 1.80                    &       0.42$_{-0.42}^{+0.68}$ 	&42.62 &    -    &  0.87 & 0.11 &  -   \\
18   & J033248.0-274233 & $1.98_{-0.10}^{+0.09}$  &       2.44$_{-0.37}^{+0.49}$ 	&43.92 &   LEX   &  0.979 & 1.66 & 0.81 \\ 
20   & J033244.5-274941 & 1.80                    &       6.51$_{-1.48}^{+1.68}$ 	&43.20 &    -    &  1.016 & 0.28 &  -   \\
21   & J033244.4-275252 & 1.80                    &       $<2.64$ 			&43.81 &  BLAGN  &  3.471 & 0.07 &  -   \\
22   & J033243.3-274915 & $1.92_{-0.12}^{+0.16}$  &       $<0.68$ 			&44.16 &  BLAGN  &  1.92 & 0.50 & 0.77 \\ 
23   & J033242.0-274400 & 1.80                    &       $<0.21$ 			&42.80 &    -    &  0.73 & 0.25 &  -   \\
24   & J033242.0-275203 & 1.80                    &       4.02$_{-4.02}^{+4.59}$ 	&44.42 &  BLAGN  &  3.610 & 0.26 &  -   \\
25$^\ast$   & J033240.9-275548 & $0.57^{+0.39}_{-0.37}$  &  $0.38^{-0.30}_{+0.30}$ 	&42.88 &   ABS   &  0.625 & 0.43 &  -   \\
26   & J033239.8-274612 & 1.80                    &       4.63$_{-1.72}^{+1.98}$ 	&43.56 &    -    &  1.65 & 0.21 &  -   \\
27   & J033239.8-274851 & 1.80                    &      47.95$_{-7.79}^{+8.61}$ 	&44.50 &   HEX   &  3.064 & 0.44 &  -   \\
28   & J033239.2-274602 & 1.80                    &       1.24$_{-0.83}^{+0.99}$ 	&43.16 &  BLAGN  &  1.216 & 0.17 &  -   \\
29   & J033239.0-275701 & $2.02_{-0.19}^{+0.16}$  &       4.47$_{-0.82}^{+0.92}$ 	&42.69 &    -    &  0.30 & 1.54 & 1.18 \\ 
31   & J033237.9-275213 & $2.22_{-0.14}^{+0.11}$  &       2.14$_{-0.63}^{+0.50}$ 	&44.17 &   HEX   &  1.603 & 0.87 & 0.83 \\ 
33   & J033236.8-274407 & $1.75_{-0.16}^{+0.11}$  &       0.12$_{-0.12}^{+0.16}$ 	&43.11 &   LEX   &  0.67 & 0.64 & 0.93 \\ 
34   & J033235.0-275512 & 1.80                    &       0.86$_{-0.40}^{+0.40}$ 	&43.00 &    -    &  0.839 & 0.29 &  -   \\
36   & J033233.1-274548 & 1.80                    &       0.38$_{-0.17}^{+0.21}$ 	&41.90 &    -    &  0.33 & 0.20 &  -   \\
37   & J033232.2-274156 & 1.80                    &       0.54$_{-0.54}^{+0.89}$ 	&42.72 &    -    &  0.96 & 0.11 &  -   \\
38   & J033230.3-274505 & $2.13_{-0.11}^{+0.12}$  &       $<0.17$ 			&43.13 &  BLAGN  &  0.74 & 0.52 & 1.27 \\ 
39   & J033230.1-274530 & $1.87_{-0.08}^{+0.08}$  &       $<0.33$ 			&43.91 &  BLAGN  &  1.22 & 0.95 & 1.19 \\ 
41   & J033227.7-274145 & 1.80                    &       7.78$_{-1.21}^{+1.32}$ 	&43.11 &   HEX   &  0.667 & 0.63 &  -   \\
42   & J033227.1-274105 & $2.09_{-0.03}^{+0.03}$  &       $0.17$ 			&44.25 &   -  &  0.73 & 6.31 & 1.25 \\ 
43   & J033226.9-274146 & 1.80                    &       2.91$_{-0.73}^{+0.81}$ 	&42.84 &   LEX   &  0.737 & 0.27 &  -   \\
44   & J033226.6-274036 & $2.44_{-0.08}^{+0.09}$  &       $<0.26$ 			&43.59 &    -  &  1.031 & 0.68 & 1.12 \\ 
45   & J033225.8-274306 & 1.80                    &      12.21$_{-2.82}^{+3.29}$ 	&44.08 &   LEX   &  2.291 & 0.33 &  -   \\
46   & J033225.3-274219 & $2.23_{-0.29}^{+0.29}$  &       0.77$_{-0.77}^{+0.90}$ 	&43.59 &  BLAGN  &  1.617 & 0.23 & 1.34 \\ 
47   & J033225.1-274101 & 1.80                    &       7.22$_{-1.80}^{+1.84}$ 	&42.86 &   LEX   &  0.733 & 0.28 &  -   \\
48   & J033224.9-275601 & 1.80                    &       5.55$_{-1.30}^{+1.48}$ 	&43.43 &    -    &  1.26 & 0.30 &  -   \\
49   & J033224.3-274127 & 1.80                    &       $<0.15$ 			&42.49 &   LEX   &  0.534 & 0.26 &  -   \\
50$^\ast$   & J033219.1-274756 & $1.01^{+0.53}_{-0.47}$    &    $0.13^{+0.32}_{-0.13}$ 	&42.45 &   ABS   &  0.67 & 0.14 &  -   \\
51   & J033217.3-275221 & $2.69_{-0.44}^{+0.28}$  &      30.14$_{-8.21}^{+5.07}$ 	&43.76 &    -  &  1.097 & 0.87 & 1.27 \\ 
52   & J033217.2-274304 & $2.03_{-0.12}^{+0.16}$  &       $<0.13$			&42.80 &  BLAGN  &  0.569 & 0.45 & 0.80 \\ 
53   & J033215.1-275128 & 1.80                    &       $<0.08$ 			&42.69 &  BLAGN  &  0.675 & 0.24 &  -   \\
54   & J033214.7-275422 & 1.80                    &      18.80$_{-5.50}^{+6.08}$ 	&44.08 &   HEX   &  2.561 & 0.25 &  -   \\
55   & J033214.1-275101 & 1.80                    &       1.86$_{-0.30}^{+0.32}$ 	&41.37 &   HEX   &  0.122 & 0.51 &  -   \\
56   & J033213.3-274241 & $1.49_{-0.15}^{+0.19}$  &       2.05$_{-0.45}^{+0.73}$ 	&43.25 &    -    &  0.605 & 1.12 & 0.93 \\ 
57   & J033213.0-275238 & 1.80                    &      18.80$_{-4.63}^{+5.21}$ 	&44.15 &   HEX   &  2.562 & 0.29 &  -   \\
58   & J033211.8-274629 & 1.80                    &       2.76$_{-0.90}^{+1.01}$ 	&42.94 &    -    &  0.92 & 0.20 &  -   \\
59   & J033211.5-275214 & $1.74_{-0.18}^{+0.18}$  &       1.81$_{-0.60}^{+0.84}$ 	&43.36 &    -    &  0.97 & 0.46 & 0.68 \\ 
60   & J033211.0-274415 & $1.92_{-0.09}^{+0.12}$  &       $<0.51$ 			&44.22 &  BLAGN  &  1.615 & 0.89 & 1.19 \\ 
61   & J033210.6-274309 & $2.02_{-0.12}^{+0.10}$  &       2.43$_{-0.94}^{+0.75}$ 	&44.28 &    -  &  2.02 & 0.83 & 0.90 \\ 
62   & J033209.5-274807 & 1.80                    &      23.20$_{-4.16}^{+4.50}$ 	&44.39 &  BLAGN  &  2.810 & 0.42 &  -   \\
63   & J033208.7-274735 & $1.93_{-0.03}^{+0.03}$  &       $<0.04$			&43.82 &  BLAGN  &  0.544 & 5.21 & 1.33 \\ 
64   & J033208.1-274658 & $1.59_{-0.17}^{+0.17}$  &       0.21$_{-0.11}^{+0.18}$ 	&41.37 &    -    &  0.13 & 0.46 & 1.44 \\ 
65   & J033204.0-275330 & 1.80                    &       0.50$_{-0.50}^{+0.57}$ 	&43.10 &    -  &  1.10 & 0.19 &  -   \\
66   & J033203.7-274604 & 1.80                    &       8.99$_{-1.16}^{+1.26}$ 	&43.22 &   LEX   &  0.574 & 1.17 &  -   \\
67   & J033202.5-274601 & $1.75_{-0.12}^{+0.13}$  &       $<0.51$ 			&44.06 &  BLAGN  &  1.616 & 0.69 & 0.76 \\ 
68   & J033201.6-274327 & $2.15_{-0.22}^{+0.28}$  &       6.87$_{-2.95}^{+2.95}$ 	&44.27 &  BLAGN  &  2.726 & 0.45 & 0.79 \\ 
70   & J033201.5-274648 & $0.88_{-0.28}^{+0.53}$  &       8.11$_{-4.06}^{+6.60}$ 	&43.78 &    -    &  1.07 & 0.96 & 1.83 \\ 
72   & J033158.3-275043 & $2.15_{-0.21}^{+0.19}$  &      10.38$_{-2.19}^{+3.10}$ 	&44.09 &    -  &  1.99 & 0.46 & 1.12 \\ 
73   & J033158.2-274834 & 1.80                    &       0.77$_{-0.30}^{+0.30}$ 	&42.93 &   LEX   &  0.734 & 0.34 &  -   \\ 
76   & J033152.6-275018 & 1.80      	       	  &      19.82$_{-3.56}^{+3.81}$ 	&44.34 &   LEX   &  2.394 & 0.53 &  -   \\ 
77   & J033301.7-274543 & 1.80                    &       $<0.25$			&42.45 &  BLAGN  &  0.622 & 0.16 &  -   \\
78   & J033230.1-274524 & $1.97_{-0.10}^{+0.11}$  &       $<0.24$			&43.48 &  BLAGN  &  0.960 & 0.62 & 1.30 \\ 
79   & J033238.1-274627 & 1.80                    &       $<1.25$			&43.41 &    -    &  1.82 & 0.12 &  -   \\
80   & J033211.0-274857 & 1.80                    &       $<0.56$ 			&43.33 &    -    &  1.70 & 0.12 &  -   \\
81   & J033226.0-274515 & 1.80                    &       5.03$_{-4.14}^{+5.32}$ 	&43.61 &    -    &  2.59 & 0.08 &  -   \\
82   & J033214.9-275104 & 1.80                    &      10.32$_{-4.99}^{+6.33}$ 	&43.37 &    -    &  1.89 & 0.10 &  -   \\
83   & J033215.0-274225 & 1.80                    &       0.29$_{-0.29}^{+1.33}$ 	&43.49 &    -    &  1.76 & 0.15 &  -   \\
84   & J033246.9-274212 & 1.80                    &       $<0.04$ 			&40.63 &   ABS   &  0.10 & 0.13 &  -   \\
85   & J033244.7-274836 & 1.80                    &       9.17$_{-5.32}^{+6.51}$ 	&43.68 &   LEX   &  2.593 & 0.10 &  -   \\
86   & J033233.9-274521 & 1.80                    &      30.09$_{-24.24}^{+34.27}$ 	&43.55 &    -    &  3.09 & 0.05 &  -   \\
89   & J033208.3-274153 & 1.80                    &       $<1.08$ 			&43.33 &  BLAGN  &  2.47 & 0.05 &  -   \\
91   & J033242.9-274703 & 1.80                    &       3.12$_{-3.12}^{+5.35}$ 	&43.95 &  BLAGN  &  3.19 & 0.11 &  -   \\
93   & J033202.4-275235 & 1.80                    &       0.64$_{-0.64}^{+1.91}$ 	&42.79 &    -    &  1.30 & 0.06 &  -   \\
95   & J033229.9-274425 & 1.80                    &       $<0.02$ 			&40.15 &   LEX   &  0.08 & 0.09 &  -   \\
96   & J033220.9-275223 & 1.80                    &       0.23$_{-0.23}^{+0.43}$ 	&41.15 &  Galaxy &  0.27 & 0.06 &  -   \\
99   & J033205.2-275356 & $1.60_{-0.20}^{+0.26}$  &       0.65$_{-0.42}^{+0.70}$ 	&43.04 &    -    &  0.79 & 0.36 & 1.53 \\ 
101  & J033255.6-274752 & 1.80                    &       $<0.51$ 			&43.19 &    -    &  1.625 & 0.09 &  -   \\
103  & J033228.9-274356 & 1.80                    &       $<0.10$ 			&41.16 &   ABS   &  0.21 & 0.10 &  -   \\
108  & J033205.8-274447 & 1.80                    &       $<0.48$ 			&42.91 &    -    &  1.56 & 0.05 &  -   \\
110  & J033258.7-274633 & 1.80                    &       $<0.57$ 			&41.85 &   LEX   &  0.622 & 0.04 &  -   \\
114$^\ast$  & J033207.7-275214 & $1.02^{+0.26}_{-0.26}$   &       $<0.07$	&43.45 &    -    &  1.72 & 0.15 &  -   \\
116  & J033230.1-274405 & 1.80                    &       $0.04$ 			&40.11 &   LEX   &  0.08 & 0.08 &  -   \\
117  & J033203.1-274450 & 1.80                    &       1.75$_{-1.75}^{+3.79}$ 	&43.69 &   HEX   &  2.573 & 0.10 &  -   \\
122  & J033257.7-274549 & 1.80                    &       1.80$_{-1.80}^{+3.21}$ 	&43.41 &    -    &  2.10 & 0.08 &  -   \\
132  & J033244.1-275456 & 1.80                    &       2.67$_{-1.89}^{+2.67}$ 	&42.45 &   LEX   &  0.908 & 0.07 &  -   \\
133  & J033202.6-274430 & 1.80                    &       3.52$_{-2.45}^{+3.35}$ 	&42.80 &    -    &  1.21 & 0.08 &  -   \\
145  & J033222.6-274604 & 1.80                    &      15.08$_{-4.20}^{+4.88}$ 	&43.50 &    -    &  1.50 & 0.23 &  -   \\
146$^\ast$  & J033247.2-275335 & $1.02^{+0.56}_{-0.30}$  &      $<0.31$ 	&43.83 &    -  &  2.67 & 0.13 &  -   \\
147  & J033246.4-274632 & 1.80                    &      25.02$_{-5.95}^{+7.87}$ 	&43.23 &    -    &  0.99 & 0.33 &  -   \\
148  & J033235.3-275319 & 1.80                    &      13.01$_{-4.19}^{+5.20}$ 	&43.56 &    -    &  1.74 & 0.18 &  -   \\
149$^\ast$  & J033212.3-274621 & $0.03^{+0.61}_{-0.64}$  &   $<0.24$ 	&42.64 &   LEX   &  1.03 & 0.08 &  -   \\
150  & J033225.3-275451 & 1.80                    &      27.79$_{-7.83}^{+10.51}$ 	&43.11 &   ABS   &  1.090 & 0.20 &  -   \\
151  & J033220.6-274733 & 1.80                    &      19.84$_{-4.48}^{+5.91}$ 	&42.74 &   LEX   &  0.604 & 0.34 &  -   \\
152  & J033259.4-274859 & 1.80                    &      21.14$_{-3.46}^{+3.91}$ 	&43.66 &    -    &  1.28 & 0.48 &  -   \\
153$^\ast$  & J033218.4-275056 & $-0.38^{+0.77}_{-0.28}$  &    $<1.4$ 	&43.68 &   HEX   &  1.536 & 0.33 &  -   \\
155  & J033208.0-274240 & 1.80                    &       1.63$_{-0.99}^{+1.37}$ 	&42.03 &   LEX   &  0.545 & 0.08 &  -   \\
156  & J033213.3-275530 & 1.80                    &      64.34$_{-16.76}^{+21.71}$ 	&43.42 &   ABS   &  1.185 & 0.33 &  -   \\
159  & J033250.4-275253 & $1.64_{-0.16}^{+0.17}$  &       9.54$_{-3.82}^{+6.68}$ 	&44.65 &    -  &  3.30 & 0.53 & 1.28 \\ 
170  & J033246.5-275414 & 1.80                    &       0.50$_{-0.50}^{+1.53}$ 	&41.98 &   ABS   &  0.664 & 0.05 &  -   \\
171  & J033235.2-274411 & 1.80                    &       $<1.36$ 			&41.89 &   LEX   &  0.84 & 0.02 &  -   \\
183  & J033234.2-275641 & 1.80         	          &       1.79$_{-1.01}^{+1.97}$ 	&40.06 &    -    &  0.08 & 0.06 &  -   \\
184  & J033248.3-275257 & 1.80                    &       7.16$_{-4.51}^{+8.99}$ 	&42.32 &   ABS   &  0.667 & 0.10 &  -   \\
185  & J033211.0-274343 & 1.80                    &       7.25$_{-4.74}^{+8.45}$ 	&42.24 &    -    &  0.93 & 0.04 &  -   \\
188  & J033222.6-274950 & 1.80                    &       7.01$_{-3.38}^{+5.13}$ 	&42.26 &   LEX   &  0.734 & 0.07 &  -   \\
189$^\ast$  & J033245.9-274213 & $0.68^{+0.86}_{-0.76}$  &     $<0.40$ 	&42.06 &    -    &  0.755 & 0.04 &  -   \\
190  & J033236.0-274100 & 1.80                    &      17.69$_{-5.43}^{+8.72}$ 	&42.79 &   LEX   &  0.733 & 0.24 &  -   \\
200  & J033255.0-274506 & 1.80                    &       2.40$_{-0.97}^{+1.12}$ 	&42.80 &    -    &  0.85 & 0.17 &  -   \\
201  & J033239.1-274440 & 1.80                    &       2.17$_{-0.83}^{+1.03}$ 	&42.49 &    -    &  0.679 & 0.15 &  -   \\
202  & J033229.9-275106 & 1.80                    &      36.85$_{-15.70}^{+20.54}$ 	&44.12 &   HEX   &  3.700 & 0.12 &  -   \\
205  & J033217.2-274137 & 1.80                    &      13.52$_{-6.52}^{+10.50}$ 	&43.06 &    -    &  1.56 & 0.08 &  -   \\
210  & J033238.4-275554 & 1.80                    &       1.72$_{-1.72}^{+4.72}$ 	&42.95 &    -    &  1.73 & 0.05 &  -   \\
218  & J033216.5-275200 & 1.80                    &       $<0.09$ 			&41.49 &    -    &  0.497 & 0.03 &  -   \\
222  & J033254.6-274501 & 1.80                    &       0.30$_{-0.30}^{+0.68}$ 	&42.97 &    -    &  1.14 & 0.13 &  -   \\
226  & J033204.5-274644 & 1.80                    &       0.11$_{-0.11}^{+1.40}$ 	&43.07 &    -    &  1.45 & 0.09 &  -   \\
227  & J033205.4-274644 & 1.80                    &      65.42$_{-16.52}^{+22.09}$ 	&43.94 &    -    &  2.18 & 0.26 &  -   \\
230  & J033153.6-274844 & 1.80                    &       1.30$_{-1.30}^{+4.54}$ 	&43.27 &  BLAGN  &  2.185 & 0.06 &  -   \\ 
236  & J033211.5-275006 & 1.80                    &       $<0.18$ 			&41.94 &    -    &  0.76 & 0.03 &  -   \\
239  & J033236.2-275127 & 1.80                    &       0.44$_{-0.44}^{+3.84}$ 	&42.55 &    -    &  1.47 & 0.03 &  -   \\
240  & J033259.2-275140 & 1.80                    &       3.70$_{-2.78}^{+3.70}$ 	&42.84 &    -    &  1.41 & 0.06 &  -   \\
241  & J033224.3-274258 & 1.80                    &       0.33$_{-0.33}^{+1.26}$ 	&41.96 &    -    &  0.70 & 0.04 &  -   \\
247  & J033234.9-275535 & 1.80                    &       0.27$_{-0.27}^{+0.80}$ 	&39.16 &   LEX   &  0.04 & 0.04 &  -   \\
248  & J033210.3-275418 & 1.80                    &       8.64$_{-4.54}^{+9.56}$ 	&42.19 &   ABS   &  0.685 & 0.07 &  -   \\
249  & J033219.5-275406 & 1.80                    &       $<0.12$ 			&42.31 &    -    &  0.96 & 0.04 &  -   \\
251  & J033207.2-275229 & 1.80                    &      27.56$_{-14.60}^{+25.30}$ 	&43.42 &    -    &  2.13 & 0.08 &  -   \\
252  & J033247.1-274346 & 1.80        	          &      16.09$_{-5.68}^{+7.65}$ 	&43.05 &   LEX   &  1.180 & 0.14 &  -   \\
253  & J033220.1-274448 & 1.80                    &      10.77$_{-2.85}^{+4.24}$ 	&42.34 &   LEX   &  0.484 & 0.23 &  -   \\
254  & J033219.9-274519 & 1.80                    &       6.37$_{-2.24}^{+4.21}$ 	&40.63 &  Galaxy &  0.10 & 0.15 &  -   \\
256  & J033243.1-274845 & 1.80                    &      33.70$_{-11.47}^{+16.03}$ 	&43.33 &    -    &  1.53 & 0.15 &  -   \\
257  & J033213.5-274857 & 1.80                    &      13.45$_{-7.47}^{+16.10}$ 	&42.13 &    -    &  0.549 & 0.10 &  -   \\
259  & J033206.2-274928 & 1.80                    &      46.86$_{-11.05}^{+13.56}$ 	&43.69 &    -    &  1.76 & 0.24 &  -   \\
260  & J033225.2-275044 & 1.80                    &      $54.90$ 			&41.78 &   LEX   &  1.043 & 0.01 &  -   \\
261  & J033157.1-275110 & 1.80                    &       2.04$_{-0.53}^{+0.66}$ 	& -  &    -    &   -   & 0.19 &  -   \\ 
263  & J033218.9-275136 & 1.80                    &      61.41$_{-54.33}^{+169.47}$ 	&43.69 &    -  &  3.660 & 0.05 &  -   \\
264  & J033229.8-275146 & 1.80                    &      27.72$_{-10.70}^{+17.49}$ 	&43.09 &   LEX   &  1.318 & 0.12 &  -   \\
265  & J033233.4-274236 & 1.80                    &      16.06$_{-3.89}^{+4.55}$ 	&43.34 &    -    &  1.22 & 0.25 &  -   \\
266  & J033214.0-274249 & 1.80                    &      38.89$_{-16.11}^{+24.49}$ 	&42.54 &   LEX   &  0.735 & 0.14 &  -   \\
505$^\ast$  & J033305.0-274732 & 1.80                    &     130$_{-63}^{+122}$ 	&43.61 &    -    &  2.26 & 0.11 &  -   \\
507  & J033300.1-274925 & 1.80                    &       3.53$_{-3.47}^{+8.18}$ 	&42.39 &    -    &  0.99 & 0.05 &  -   \\
508  & J033251.8-275214 & 1.80                    &      57.52$_{-27.38}^{+47.57}$ 	&43.58 &    -  &  2.50 & 0.08 &  -   \\
510  & J033238.8-275122 & 1.80                    &      43.75$_{-32.60}^{+92.51}$ 	&43.34 &    -    &  2.51 & 0.05 &  -   \\
511  & J033236.6-274631 & 1.80                    &       $<1.41$ 			&41.68 &    -    &  0.767 & 0.02 &  -   \\
512  & J033234.5-274351 & 1.80                    &       $<0.82$ 			&41.72 &   LEX   &  0.665 & 0.03 &  -   \\
513  & J033234.1-274900 & 1.80                    &      $<12.82$ 			&43.31 &    -    &  3.56 & 0.02 &  -   \\
514  & J033233.6-274312 & 1.80                    &       $<0.42$ 			&39.85 &   ABS   &  0.10 & 0.02 &  -   \\
515  & J033232.2-274652 & 1.80                    &      50.29$_{-23.98}^{+41.21}$ 	&43.51 &    -    &  2.28 & 0.09 &  -   \\
516  & J033231.4-274727 & 1.80                    &       1.13$_{-1.13}^{+2.18}$ 	&41.92 &   LEX   &  0.665 & 0.04 &  -   \\
518  & J033226.9-274605 & 1.80                    &       2.47$_{-2.11}^{+4.33}$ 	&42.10 &    -    &  0.84 & 0.04 &  -   \\
519  & J033225.9-275508 & 1.80                    &       $<1.96$ 			&42.23 &   LEX   &  1.034 & 0.03 &  -   \\
521  & J033222.8-275225 & 1.80                    &       0.15$_{-0.15}^{+0.39}$ 	&40.34 &   LEX   &  0.131 & 0.04 &  -   \\
522  & J033221.5-275550 & 1.80                    &       2.04$_{-2.04}^{+6.12}$ 	&43.46 &    -    &  2.57 & 0.06 &  -   \\
523  & J033220.4-274229 & 1.80                    &      10.70$_{-6.32}^{+9.21}$ 	&42.70 &    -    &  1.32 & 0.05 &  -   \\
524  & J033220.0-274243 & 1.80                    &      16.13$_{-9.93}^{+12.66}$ 	&43.40 &    -    &  2.36 & 0.06 &  -   \\
525  & J033219.9-274123 & 1.80                    &       0.23$_{-0.23}^{+0.54}$ 	&40.84 &   LEX   &  0.229 & 0.04 &  -   \\
526  & J033218.8-274413 & 1.80                    &       8.51$_{-8.51}^{+16.30}$ 	&42.18 &    -    &  0.96 & 0.03 &  -   \\
527  & J033218.6-275414 & 1.80                    &       7.29$_{-7.29}^{+26.44}$ 	&43.87 &    -    &  4.49 & 0.04 &  -   \\
528  & J033217.1-275402 & 1.80                    &       $<0.32$ 			&42.30 &    -    &  1.43 & 0.02 &  -   \\
529  & J033216.3-275525 & 1.80                    &       4.02$_{-4.02}^{+5.04}$ 	&42.10 &    -    &  0.73 & 0.05 &  -   \\
531$^\ast$  & J033214.5-275111 & $-0.96^{+1.08}_{-2.04}$  &       $<3.2$ 			&42.49 &   HEX   &  1.544 & 0.02 &  -   \\
532  & J033214.2-274231 & 1.80                    &       0.18$_{-0.18}^{+1.82}$ 	&42.12 &    -    &  0.95 & 0.03 &  -   \\
534  & J033212.2-274530 & 1.80     	          &      13.09$_{-4.98}^{+10.40}$ 	&42.46 &   LEX   &  0.676 & 0.14 &  -   \\
535$^\ast$  & J033211.5-274650 & $0.59^{+0.52}_{-0.50}$ &    $<0.29$	&42.12 &   LEX   &  0.575 & 0.09 &  -   \\
536  & J033210.9-274235 & 1.80                    &       0.79$_{-0.79}^{+1.72}$ 	&41.40 &    -    &  0.42 & 0.04 &  -   \\
537  & J033209.9-275016 & 1.80                    &       3.90$_{-3.90}^{+8.04}$ 	&42.68 &    -    &  1.54 & 0.03 &  -   \\
538$^\ast$  & J033208.6-274649 & $-1.1^{+1.7}_{-1.0}$    &   $<1.0$ 	&41.62 &   LEX   &  0.310 & 0.12 &  -   \\
540  & J033202.7-275053 & 1.80                    &       3.52$_{-3.52}^{+13.12}$ 	&42.33 &    -    &  1.25 & 0.02 &  -   \\
541  & J033159.7-274949 & 1.80                    &       7.18$_{-5.15}^{+7.49}$ 	&43.07 &    -    &  1.82 & 0.05 &  -   \\ 
543  & J033157.0-275102 & 1.80                    &       3.55$_{-3.55}^{+12.21}$ 	&42.87 &    -    &  1.81 & 0.03 &  -   \\ 
544  & J033154.6-275104 & 1.80                    &      20.35$_{-10.92}^{+15.64}$ 	&43.42 &    -    &  2.36 & 0.07 &  -   \\ 
597  & J033251.4-275544 & 1.80                    &      25.01$_{-23.32}^{+69.25}$ 	&43.01 &    -    &  2.32 & 0.03 &  -   \\
598  & J033224.7-275413 & 1.80                    &       4.13$_{-4.13}^{+11.67}$ 	&41.71 &   ABS   &  0.617 & 0.03 &  -   \\
599$^\ast$  & J033229.9-275329 & 1.80                    &     100.0$_{-48}^{+65}$ 	&43.58 &    -    &  2.84 & 0.06 &  -   \\
600$^\ast$  & J033213.9-274526 & 1.80                    &     380.0$_{-180}^{+180}$ 	&44.43 &   LEX   &  1.33 & 0.26 &  -   \\ 
602  & J033222.0-274657 & 1.80                    &      97.81$_{-58.69}^{+106.02}$ 	&42.24 &   ABS   &  0.668 & 0.08 &  -   \\ 
603  & J033257.8-274711 & 1.80                    &      23.80$_{-23.80}^{+54.07}$ 	&42.82 &    -    &  2.04 & 0.02 &  -   \\
605$^\ast$  & J033239.2-274833 & 1.80                    &    $2700^{+3800}_{-1700}$ 			&44.95 &    -    &  4.29 & 0.12 &  -   \\
606  & J033225.0-275009 & 1.80                    &      20.11$_{-9.79}^{+17.79}$ 	&42.68 &    -    &  1.037 & 0.08 &  -   \\
607  & J033159.7-275020 & 1.80                    &      16.32$_{-6.99}^{+14.49}$ 	&42.46 &    -    &  0.74 & 0.11 &  -   \\ 
609$^\ast$  & J033236.2-275037 & 1.80             &     250$_{-170}^{+400}$ 	&43.12 &    -    &  1.86 & 0.06 &  -   \\ 
610$^\ast$  & J033219.9-275159 & 1.80                    &    520$_{-280}^{+580}$	&44.18 &    -    &  2.04 & 0.10 &  -   \\ 
611  & J033241.7-274328 & 1.80                    &      30.99$_{-14.91}^{+32.82}$ 	&42.64 &    -    &  0.979 & 0.09 &  -   \\
612  & J033221.4-274231 & 1.80                    &      35.72$_{-13.50}^{+26.07}$ 	&42.93 &    -    &  1.10 & 0.13 &  -   \\
615  & J033201.3-275052 & 1.80                    &      10.42$_{-10.42}^{+25.32}$ 	&42.03 &   LEX   &  0.759 & 0.04 &  -   \\
632  & J033233.5-275228 & 1.80                    &      33.91$_{-23.67}^{+60.75}$ 	&42.91 &    -    &  1.57 & 0.05 &  -   \\
634$^\ast$  & J033251.5-274746 & 1.80                    &  $460^{+1500}_{-460}$	&42.72 &    -    &  1.40 & 0.05 &  -   \\
635  & J033216.9-275009 & 1.80                    &       1.50$_{-1.50}^{+3.55}$ 	&41.72 &    -    &  0.729 & 0.02 &  -   \\
637  & J033225.7-274332 & 1.80                    &      33.64$_{-31.62}^{+208.04}$ 	&42.12 &    -    &  0.76 & 0.05 &  -   \\
638  & J033230.0-274302 & 1.80                    &       4.83$_{-4.83}^{+407.27}$ 	&42.40 &    -    &  1.39 & 0.02 &  -   \\
642$^\ast$  & J033215.3-274159 & 1.80             &     130$_{-110}^{+\infty}$ 	&43.32 &   HEX   &  2.402 & 0.05 &  -   \\ 
\end{longtable}

\begin{list}{}{}
\item$^1$Source name from Giacconni et al. (2002)
\item{$^2$Best fit model power-law photon index} 
\item{$^3$Rest-frame column density in units of $10^{22}$ (\cunits)}
\item{$^4$logarithm of 2-10 keV intrinsic luminosity (\lunits)} 
\item{$^5$Optical Spectroscopic classification as given in Szokoly et al. (2004)}
\item{$^6$Redshift from either Szokoly et al. (2004) (spectroscopic) or Zheng 
 et al. 2004 (photometric)}
\item{$^7$observed 2-10keV X-ray flux in units of $10^{-14}$ \funits}
\item{$^9$Reduced $\chi^2$} 
\item{$^\ast$ Compton thick sources}
\end{list}

\end{document}